# Re-entrant unconventional superconductivity induced by rare-earth substitution in $Nd_{1-x}Eu_xNiO_2$ thin films

Dung Vu[1], Hangoo Lee[2], Daniele Nicoletti[2], Wenzheng Wei[1], Zheting Jin[3], Dmitry V. Chichinadze[4], Michele Buzzi[2], Wenxin Li[1], Xinhao Yang[1], Rongting Wu[1], Christopher A. Mizzi[5], Tiema Qian[5], Boris Maiorov[5], Alexey Suslov[4], Yu He[1,3], Cyprian Lewandowski[4,6], Sohrab Ismail-Beigi[1,3,7], Frederick Walker[1], Andrea Cavalleri[2,8], Charles Ahn[1,3,7]

[1]*Department of Applied Physics, Yale University, New Haven, CT 06520, USA.*

[2]*Max Planck Institute for the Structure and Dynamics of Matter, 22761 Hamburg, Germany*

[3]*Department of Physics, Yale University, New Haven, CT 06520, USA.*

[4]*National High Magnetic Field Laboratory, Tallahassee, FL 32310, USA*

[5] *National High Magnetic Field Laboratory, Los Alamos National Laboratory, Los Alamos, NM 87545, USA*

[6]*Department of Physics, Florida State University, Tallahassee, Florida 32306, USA*

[7]*Department of Mechanical Engineering and Materials Science, Yale University, New Haven, CT 06520, USA.*

[8]*Department of Physics, Clarendon Laboratory, University of Oxford, Oxford OX1 3PU, United Kingdom*

## Abstract

High temperature superconductivity is typically associated with strong coupling and a large superconducting gap, yet these characteristics have not been demonstrated in the nickelates. Here, we provide experimental evidence that Eu substitution in the spacer layer of $Nd_{1-x}Eu_xNiO_2$ (NENO) thin films enhances the superconducting gap, driving the system toward a strong-coupling regime. This is accompanied by a magnetic-exchange-driven magnetic-field-enhanced superconductivity. We investigate the upper critical magnetic field, $H_{c2}$, and superconducting gap of superconducting NENO thin films with x=0.2 to 0.35. Magnetoresistance measurements reveal magnetic-field-enhanced superconductivity in NENO films. We interpret this phenomenon as a result of interaction between magnetic Eu ions and superconducting states in the Ni $d_{x2-y2}$ orbital. The upper critical magnetic field strongly violates the weak-coupling Pauli limit. Infrared spectroscopy confirms a large gap-to-$T_c$ ratio $2\Delta/k_BT_c \simeq 5-6$, indicating a stronger coupling pairing mechanism in NENO relative to the Sr-doped $NdNiO_2$. The substitution of Eu in the rare-earth layer causes pronounced modifications of the superconducting gap and magnetic interactions in Nd-based nickelates, opening new pathways to engineer high-$T_c$ superconductivity in infinite-layer nickelates.

## Main

The discovery of superconductivity in layered nickelates (A, B)$NiO_2$ (A=La, Pr, Nd, Sm; B=Sr, Ca, Eu) has brought about new perspectives in the study of high-temperature superconductors (*1-6*). While they share some common features with the copper-based high $T_c$ superconductors (the cuprates), such as stacked two-dimensional metal oxide layers and the main contribution to low-energy bands from the $d_{x2-y2}$ orbital, there are significant differences between these systems (*7-11*). Superconductivity in the cuprates is believed to be strongly coupled and driven by strong electron-electron interactions arising from the Cu $3d_{x2-y2}$ band and spin fluctuations(*12*). In contrast, there has not been direct evidence for strong coupling in known nickelate superconductors (*13, 14*). Additionally, while the spacer layer in the cuprates plays little role in the electronic structure near the Fermi level (*7, 15*), there is a considerable variation in superconducting properties within the $RENiO_2$ family (RE=rare earth), depending on the RE-site ion (**Figure 1a**).

It is theoretically suggested that the spin structures, exchange interactions, and Fermi surfaces in nickelates strongly depend on the types of RE elements (*16-19*). Such changes are manifested in several experimental studies. The upper critical field $H_{c2}$ of La-based nickelate $La_{1-x}Sr_xNiO_2$ (LSNO) thin films shows anisotropy between out-of-plane (**H**||c) and in-plane (**H**||ab) magnetic fields and a violation of the Pauli limit (*16, 20-22*), which is a bound on the upper critical magnetic field of a weakly-coupled singlet superconductor. In contrast, the isotropic, Pauli-limited $H_{c2}$ in $Nd_{0.775}Sr_{0.225}NiO_2$ (NSNO) (*23*) makes NSNO more similar to the iron-based high-$T_c$ superconductors that have multigap pairing (*24, 25*). London penetration depth studies in nickelates indicate that while the superconducting gap in LSNO is nodal and anisotropic, the superconducting gap in NSNO is fully gapped with differing interpretations regarding the symmetry of the gap (*25, 26*). Superconducting gap measurements on NSNO reveal variations in gap values and symmetries, with evidence for multiple gaps (*27*). Most data on NSNO suggest a superconducting gap consistent with BCS-like weak coupling, with gap-to-$T_c$ ratios of $2\Delta/k_BT_c \approx 3-3.4$ (*13, 14*). These experimental observations highlight the importance of RE chemistry in modifying the superconducting gap in nickelates.

Motivated by these findings and our initial observation of anomalously high $H_{c2}$ in $Nd_{1-x}Eu_xNiO_2$ (NENO) thin films with measurements up to 14 T (*5*), we present here comprehensive $H_{c2}(T)$ measurements in magnetic field up to 60 T, which are essential to reveal the re-entrant superconductivity behavior in NENO. We identify the microscopic origin of this effect through density functional theory and modelling of the temperature dependence of $H_{c2}(T)$. Our key findings demonstrate that Eu substitutions in the RE layer not only provide charge doping but also introduce exchange coupling between Eu $4f$ magnetic moments and Ni $3d_{x^2-y^2}$ electrons, forming a rare occurrence of the Jaccarino–Peter effect in a thin film. Optical spectroscopy reveals a large superconducting gap ($2\Delta \approx 75$ cm$^{-1}$), providing direct evidence for strong-coupling superconductivity and explaining the observed Pauli limit violation of $H_{c2}$. Together, these results establish NENO as a strongly coupled unconventional superconductor with pronounced RE magnetism effect, in sharp contrast to the weak-coupling behavior of Sr-doped $NdNiO_2$. We demonstrate that Eu substitution offers a viable method to tune pairing strength and magnetic interactions in infinite layer nickelates.

## Magnetic-field-enhanced superconductivity and Pauli limit violation

In conventional superconductors, magnetic fields typically suppress superconductivity through vortex formation or Pauli paramagnetic de-pairing (*28*). The former leads to the loss of superconducting coherence; the latter raises the energy difference between spin-up and spin-down electrons, making the formation of Cooper pairs energetically unfavorable. In a singlet superconductor, the paramagnetic effect sets an upper bound on the upper critical magnetic field, $H_{c2}^P = \sqrt{2}\Delta/g\mu_B$, where $\mu_B$ is the Bohr magneton and g is the Lande factor. For weakly-coupled superconductors, using $2\Delta=3.528k_BT_c$ (*28*), and $g \approx 2$, this simplifies to the nominal Pauli limit, $H_{c2}^P = 1.86\ T_c$(K), at $T$=0 K. Superconductors that not only withstand high magnetic fields but also have superconductivity stabilized by magnetism are highly unconventional. Here, we show that superconducting NENO thin films exhibit magnetic-field-enhanced superconductivity and attribute this phenomenon to magnetism of the Eu dopant.

We start by examining the magnetoresistance (MR) of three $Nd_{1-x}Eu_xNiO_2$ thin films which were measured with magnetic fields up to 41 T for both **H**||c and **H**||ab at the National High Magnetic Field Laboratory (NHMFL). The films are grown using the same procedure described in Ref (*5*) and consist of 6 nm thick NENO films with x=0.20, 0.22 and 0.35 on $(LaAlO_3)_{0.3}(Sr_2TaAlO_6)_{0.7}$ (LSAT) substrates with a 1 nm $Al_2O_3$ capping layer (**Figure 1b**). **Figure 2a-f** shows color plots representing temperature and applied magnetic field dependences of the resistance *R(T, H)*, interpolated from the raw *R(H)* curves taken at fixed temperatures. The black curves tracking the evolution of superconducting transition temperature taken by 50% *Rn* (*i.e.* $T_{c,50\%Rn}$), where *Rn* is the resistance in the normal state, are overlaid in **Figure 2a-f**, giving us the temperature dependence of $H_{c2}$, $H_{c2}(T)$.

The applied magnetic fields suppress superconductivity in NENO in a highly non-monotonic and distinctive manner when compared to data reported in NSNO (*23*). First, *R(T, H)* for **H**||c and **H**||ab are anisotropic, and the superconducting state persists to a high field. The $H_{c2}(T)$ curves under **H**||c show broadening of the superconducting phase transition, suggesting the presence of a vortex liquid phase (*29, 30*). In contrast, under **H**||ab, there is no apparent broadening, suggesting the dominance of paramagnetic depairing and a negligible role of vortex dynamics and thermal fluctuation. At the same temperature where the paramagnetic effect is dominant, $H_{c2,\boldsymbol{H}||ab}$ is much larger than $H_{c2,\boldsymbol{H}||c}$. $H_{c2}(T)$ shows that superconductivity in all samples reported here persists to the highest applied magnetic fields and strongly violates the Pauli limit for both **H**||c and **H**||ab. This distinguishes NENO from NSNO (*23*) and brings it closer to LSNO (*16, 20-22*). Dissipation in the superconducting phase is observed at low field as an increase in *R(H)*, shown in **Figure 2g-h**. This dissipation is then suppressed over a wide range of magnetic fields centered around 20 T. *R(H)* shows a maximum at low field and a minimum at ~ 20 T that reaches zero resistance at low temperatures. This anomalous behavior is often described as re-entrant superconductivity (*31-36*) and has not been reported in the nickelates.

Due to the anomaly in *R(H)*, $H_{c2}(T)$ shows anomalous inflection points and an enhancement of superconductivity, manifested as an increased $dH_{c2}/dT$ in **H**||c and an increased $T_c$ in **H**||ab. Similar field and temperature dependence of MR, especially a magnetic-field-enhanced superconductivity (*37-39*) have been explained by proximal magnetism, which is the Jaccarino-Peter (J-P) effect (*40*). As depicted in **Figure 3a**, the J-P effect involves an exchange field $\mathbf{H_J}$ that acts on the conducting electrons, and comes from the interaction $\widehat{H}_{ex} = \sum_{ik} J_{ik}\boldsymbol{S_i}\cdot\boldsymbol{s_k}$ between local magnetic moments $\boldsymbol{S_i}$ and the conduction electron spin density $\boldsymbol{s_k}$, where i and k label nearest neighbor RE and Ni orbitals. The J-P effect requires an antiferromagnetic coupling with a positive sign of the exchange coupling. As the applied magnetic field **H** polarizes the localized moments **S**, the antiferromagnetic exchange field $\mathbf{H_J}$ acting on $\boldsymbol{s_k}$ is $H_J = \sum_i J_{ik}\boldsymbol{S_i}$, which opposes the external field **H** and reduces the total magnetic field on the carriers in the nickel layers, $\mathbf{H_T}=\mathbf{H}+\mathbf{H_J}$, enabling superconductivity at higher applied fields. As shown in **Figure 3b**, $|\mathbf{H_T}|$ exhibits a local maximum and a compensation point at $H$~$H_{J0}$, where $H_{J0}$ is the saturated exchange field when the local moments are polarized. Depending on the magnitude of the Pauli limited $H_{c2}^P$ compared to $H_{J0}$, one can expect re-entrant superconductivity ($H_{J0}>H_{c2}$) or magnetic-field-enhanced superconductivity ($H_{J0}<H_{c2}$). Previous work revealed mixed $Eu^{2+}$ and $Eu^{3+}$ valences on the RE-site cation with a large $Eu^{2+}$ ($J$=7/2) local moment in NENO (*5*). Qualitatively, the $H_{c2}(T)$ data strongly resembles the case in which the Pauli limited $H_{c2}>H_{J0}$. The temperature dependence of the maxima and minima of *R(H)* curves under **H**||ab follows that of $|\mathbf{H_T}|$ calculated using a Brillouin function $B_J$(T, H) describing polarization of paramagnetic $Eu^{2+}$ moment in an applied magnetic field (see **Supplemental Information**). We posit that the nonmonotonic behavior in NENO is a result of a large critical field $H_{c2}$ and a J-P-type mechanism due to the exchange field from the Eu ions, as explained below.

We performed density functional theory (DFT) calculation to estimate the exchange coefficient $J_{ik}$ between different orbitals of the RE ions and Ni ions. The results (see **Supplemental Information**) show non-negligible exchange fields $\mathbf{H_J}$ on Ni orbitals induced by different Eu 4f-orbitals. We ignore the contributions from Eu 5d-orbitals because their local moments are only about $0.1\mu_B$, much smaller than the 4f-orbitals (about $6.5\mu_B$). Eu ions couple antiferromagnetically to Ni $3d_{x2-y2}$ and Ni $4p_z$ while they couple ferromagnetically to Ni $3d_{z2}$. Interestingly, Eu provides a 21.3 T antiferromagnetic exchange field on Ni $3d_{x2-y2}$, closely matching the compensation field seen in our experiments. The same calculation was done for Nd and Sr dopants, and we find negligible exchange fields on all Ni orbitals (<1T for Nd, and <0.1T for Sr). The calculation results highlight the key contribution from the Eu dopant to the field compensation mechanism and also reveal which Ni orbitals are involved. From the calculations, we can infer that in NENO, electrons in the Ni $3d_{x2-y2}$ orbital are responsible for superconductivity. Since computed exchange fields for both Ni $3d_{z2}$ and Ni $4p_z$ are too large (115 T and 128 T, respectively) compared to the applied field strengths, we expect suppressed contribution to the J-P effect from these orbitals, and the large exchange fields would likely reduce contribution from Ni $3d_{z2}$ and Ni $4p_z$ electrons to superconductivity.

Considering the J-P effect, we modeled $H_{c2}(T)$ using two approaches. First, we employ a Ginzburg-Landau (G-L) approach, which is valid for both weak- and strong-coupling superconductors in the vicinity of the superconducting phase transition and relatively low magnetic field (see **Supplemental Information**). The G-L-based model we developed fits both the **H**||c and **H**||ab data up to 41 T for all samples, capturing the non-monotonic behavior of $H_{c2}(T)$ shown in Figure 2a-f. The extracted $H_{J0}$ ~17-23 T from these fits matches the DFT-calculated $H_{J0}$ of 21.3 T. **Figure 3c** shows extended $H_{c2}(T)$ data for the x=0.22 sample measured in pulsed magnetic fields up to 60 T for both **H**||c and **H**||ab in the pulsed field facility of NHMFL at Los Alamos National Laboratory. The G-L-based model continues to fit the data well up to ~50 T. To describe the data using a conventional modeling technique, $H_{c2}(T)$ is fitted to the Fischer formula (*41*), which is used to analyze J-P effect in organic and Chevrel phase superconductors (*42, 43*). The formula is based on the microscopic Werthamer–Helfand–Hohenberg theory (*44-47*). In our case, the Fischer formula fits $H_{c2}(T)$ data up to 41 T and only noticeably deviates from the data and G-L fits at higher magnetic fields (see **Supplemental Information**). At lower temperatures, neither model captures an upturn of $H_{c2}(T)$ above both fits (**Figure 3c**). An upturn in $H_{c2}(T)$ at low temperature near 0 K, also observed in NSNO and LSNO, could be related to multiband superconductivity or an emergence of an unconventional electronic state (*23, 48-50*). While a detailed analysis of this feature lies beyond the focus of our work, we propose that if higher order terms in powers of magnetic field were kept in the G-L expansion, they could account for the deviation from the parabolic profile shown in **Figure 3c.** Overall, the $H_{c2}(T)$ models support a J-P-type mechanism in NENO and suggest a strong-coupling scenario.

## Evidence for enhanced coupling strength from superconducting gap measurements

As expected from the J-P effect, $H_{c2,0K}$ in NENO is enhanced by the exchange field $\boldsymbol{H_J}$, with an enhancement as large as $H_{J0}$. This partially explains the much higher $H_{c2}$ in NENO when compared to other infinite layer nickelates. Without experimental $H_{c2}(T)$ data near $T$=0 K, we resort to estimating the intrinsic Pauli limit violation using an extrapolation to $T$=0 K of the modified G-L fit with **H**||ab (which underestimates the experimental data) and then subtracting the enhancement from the J-P effect. This estimation yields a Pauli limit violation ratio of 2.3-3.4 for NENO. Previous studies on Pauli limit violation in La-nickelates have discussed that mechanisms like finite momentum pairing and strong spin-orbit coupling are unlikely explanations (*20-22*), while spin-triplet superconductivity remains as a speculative mechanism. Strong coupling in nickelates has been proposed but it has not yet been thoroughly investigated experimentally. Using the electron spin $g$-factor $g$=2, we approximate a gap-to-$T_c$ ratio $\frac{2\Delta}{k_B T_c} =$ $\frac{H_{c2,0K}-H_{J0}}{T_c}\frac{\sqrt{2}g\mu_B}{k_B}$ of 12.1, 8.0 and 8.4 for samples with x=0.20, 0.22 and 0.35, respectively. This doping dependence is reminiscent of similar phenomenology observed in the strongly coupled hole-doped cuprates (*51-53*). This underscores potentially common aspects of unconventional superconductivity in these two layered oxide superconductors. To evaluate this possibility, we investigated the optical response of NENO by Fourier transform spectroscopy over the entire far-infrared range to determine its superconducting gap.

Reflectivity spectra from a NENO film with x=0.30 grown on LSAT ($T_{c,50\%Rn}$ at ~ 18 K) and a bare LSAT substrate were measured for different temperatures. Small changes in reflection (*54*) from the 6-nm thick NENO films were detected by dividing the data measured in the superconducting state by that taken at $T$ = 22.5 K > $T_c$ (see **Methods**). As shown in **Figure 4a**, when cooling below $T_c$ the normalized reflectivity develops a prominent dip around 75 $\mathrm{cm}^{-1}$and a sharp upturn at lower frequencies, which is specific to superconducting NENO and absent on the substrate alone. From the temperature-dependent ratios in **Figure 4b**, we find that this feature, which we associate with the opening of the superconducting gap, is no longer resolved above 15 K, flattening out completely at $T$ = 20 K. We also observed a peak centered around 170 $\mathrm{cm}^{-1}$, which we attribute to a spurious effect caused by a change in multilayer reflectivity at the LSAT phonon frequency due to the opening of the superconducting gap in NENO.

We fitted these normalized reflectivities by considering the sample as a multilayer stack consisting of an $Al_2O_3$ capping layer, NENO film, and LSAT substrate (see **Figure 1b**). Since the weakly frequency-dependent optical properties of $Al_2O_3$ are known (59) and those of the bare LSAT substrate were directly measured alongside NENO (see **Supplemental Information**), the multilayer reflectivity could be computed by modeling only the response of superconducting NENO. We started from a Mattis-Bardeen model (see **Supplemental Information**), which describes the optical properties of a superconductor (*55*) by assuming an *s*-wave gap in the dirty limit (*i.e.*, $2\Delta \ll \Gamma$, where $\Gamma$ is the normal carrier scattering rate). This model yielded the measured normalized reflectivity, using only the superconducting gap ($2\Delta$) and the normal state conductivity ($\sigma_0$) of NENO as free parameters. As shown in **Figure 4a**, the best fit for the experimental data was obtained for $2\Delta \simeq 75\ \mathrm{cm}^{-1}$ and a normal state DC conductivity of 100 $\Omega^{-1}\mathrm{cm}^{-1}$.

Note that the fitted value of $\sigma_0$ is more than one order of magnitude smaller than the 3000 $\Omega^{-1}\mathrm{cm}^{-1}$ obtained independently from electrical transport. To improve on the fit, we next considered that uncondensed quasiparticles are likely to persist down to zero temperature due to nodal excitations if the gap symmetry is lower than *s*, an effect not included in the Mattis-Bardeen fit above. We tested this hypothesis by adapting the fit to a nodal *d*-wave superconductor in the dirty limit and with a residual metallic Drude peak (*56*) (see **Supplemental Information**). This model assumes that a part of the spectral weight at low frequencies arises from uncondensed quasiparticles, a mechanism that has been used to explain missing superfluid density in cuprates (*56*). As shown in **Figure 4a**, this model fits the data with the same value of the superconducting gap ($2\Delta \simeq 75\ \mathrm{cm}^{-1}$) *and* an improved estimate of the normal-state conductivity (~500 $\Omega^{-1}\mathrm{cm}^{-1}$). In addition, it fits well the temperature-dependent spectra (see **Figure 4b**) by letting the contribution of uncondensed quasiparticles increase with $T$.

Although the model above broadly captures the normal state conductivity, the remaining difference could be adapted by assuming that the superconductor is spatially inhomogeneous, with superconducting and normal state coexisting in different regions of the sample. As shown in the **Supplemental Information**, all spectra could be fitted with the correct value of the conductivity of the normal state of $\sigma_0 \simeq 3000\ \Omega^{-1}\mathrm{cm}^{-1}$ by tuning the filling fraction between normal and superconducting regions (*57*). In spite of possible inclusion of normal regions, the observation of robust zero resistance at high magnetic fields, bolstered with a high critical current density and a large diamagnetic screening signal from mutual inductance measurement, demonstrates the bulk nature of superconductivity in our samples (see **Supplemental Information**). While this strongly suggests d-wave pairing through the persistence of low-frequency spectral weight inconsistent with fully-gapped s-wave superconductivity, momentum-resolved gap measurement techniques are expected to add clarity to the nature of the superconducting gap.

In the aggregate, regardless of the model used, the gap amplitude $2\Delta$ is consistently estimated as $2\Delta \simeq 75\ \mathrm{cm}^{-1}$ (9.3 meV). This value is about two times greater than that reported for NSNO (*13, 14*), resulting in a gap-to-$T_c$ ratio of $2\Delta/k_BT_c \simeq 5-6$, in the same range as reported for the hole doped cuprates in the overdoped range (*51-53*). The large gap-to-$T_c$ ratio may plausibly explain the unusually high upper critical field reported here in NENO and in La-based nickelates (*20, 22*). While the measured $2\Delta/k_BT_c$ does not fully align with the ratio estimated from the Pauli limit violation, the discrepancy likely arises from the distinction between zero-field FTIR measurements and high-field transport measurements, suggesting that the superconducting gap may be magnetic-field-dependent due to interactions of Eu magnetic moments with superconductivity.

The large difference in pairing strength between NSNO and NENO demonstrates that Eu substitution in the RE layer can be an effective method for modifying the superconducting gap size in the Nd-based nickelates. The presence of Eu magnetic moments and Eu-Ni exchange interactions may indirectly influence the Ni-Ni magnetic exchange interactions (*58*), and facilitate strong coupling, while Nd-Ni and Sr-Ni exchange interactions are negligible (smaller than 0.1meV based on our DFT calculations). Furthermore, the smaller ionic radius of $Eu^{2+}/Eu^{3+}$ relative to $Sr^{2+}$ is suggested to reduce the $NiO_2$ plane separation and strengthen the pairing interaction through dimensionality control (*59-61*), with the observation of a $T_c$ above 35K in the smaller RE size infinite layer nickelate $Sm_{1-x-y-z}Ca_xSr_yEu_zNiO_2$ (*6*). The reported c-axis lattice constant in NENO is smaller by 0.05Å compared to NSNO (*5, 62*), which may contribute to the observed increase in pairing strength in NENO.

# Conclusions

We show that NENO thin films exhibit magnetic-field-enhanced superconductivity driven by a Jaccarino-Peter-type effect, in which the exchange fields from local magnetic moments of Eu ions reduce the effective field acting on the superconducting states in the Ni $d_{x2-y2}$ orbital. This behavior is the first indication of a strong interaction between Eu dopants and the superconducting state. The anomalously high $H_{c2}$ exceeding the Pauli limit, in conjunction with a large $2\Delta/k_BT_c \simeq 5-6$ measured by infrared spectroscopy, confirms a strong coupling superconductivity mechanism in the Eu doped $NdNiO_2$, in stark contrast to the weak coupling mechanism suggested for the Sr doped $NdNiO_2$. The reflectivity data in NENO fits well to a dirty node-line d-wave pairing model. These observations reveal a pairing energy scale that is similar to values in the strongly coupled, hole-doped cuprates. The distinct superconducting characteristics of NENO compared to NSNO highlight the potential of dopants like Eu to tune the electronic structure in infinite-layer nickelates. Specifically, Eu magnetic moments can be manipulated to control superconducting states, and Eu substitution in the rare-earth layer offers a new dimension to tune superconducting pairing strength, an essential factor in understanding high-$T_c$ nickelates (*6*).

Acknowledgments

This work was supported by U.S. Department of Energy, Office of Science, Office of Basic Energy Sciences, under award number DE-SC0019211. This work involves the use of resources from the Yale Materials Characterization Core. A portion of this work was performed at the National High Magnetic Field Laboratory, which is supported by National Science Foundation Cooperative Agreement No. DMR-2128556 and the State of Florida and the U.S. Department of Energy. D.V.C. acknowledges financial support from the National High Magnetic Field Laboratory through a Dirac Fellowship. C.L. was supported by start-up funds from Florida State University and the National High Magnetic Field Laboratory. W. L. was partly supported by the James Kouvel Fellowship. Y. H. acknowledges support from the U.S. Air Force Office of Scientific Research under Award No. FA9550-24-1-0048. Computational studies in this work were supported by Grant No. NSF DMR 2237469, by NSF ACCESS supercomputing resources via allocation TG- MCA08X007, and by the guidance and use of research computing infrastructure at the Yale Center for Research Computing. Work in Hamburg was supported by the Deutsche Forschungsgemeinschaft (DFG) by the Cluster of Excellence CUI: Advancing Imaging of Matter (EXC 2056, project ID 390715994). We thank Dr. Danilo Ratkovski for supporting our magnet time at the National High Magnetic Field Laboratory.

# Methods

## Thin film growth and characterization

Thin films of $Nd_{1-x}Eu_xNiO_3$ are grown to a thickness of 18-21 unit cells (uc) on as-received commercial 5×5 mm LSAT (001) substrates (from CrysTec) using molecular beam epitaxy (MBE). The substrates are cleaned at 605°C by using an activated oxygen source at a chamber pressure of $8\times10^{-6}$ Torr for 15 minutes prior to thin film synthesis. The thin films are grown under the same conditions as for the plasma cleaning process. After growth, the films are cooled in activated RF plasma oxygen to 150°C to eliminate oxygen vacancies. The thin film $Nd_{1-x}Eu_xNiO_3$ is kept in the chamber for *in-situ* reduction using metallic Al in a process as described in Ref. (*63*).

θ-2θ scans (see **Supplemental Information**) of the films around the (002) pseudo-cubic peak show a lattice constant corresponding to 112 phase and finite size oscillations visible in all the curves, indicating the high crystallographic quality of the films. X-ray diffraction data shows the c-axis lattice constant of 3.31 Å in NENO with small doping dependence.

## Transport measurement at high magnetic fields

For angular magnetoresistance measurements, the samples were contacted using wire-bonded aluminum wires in a Hall bar geometry. The high-field measurements were conducted at NHMFL DC Field Facility in Tallahassee, with 35 T and 41 T resistive magnets and temperature from 2 K to 21 K using the He-3 insert. The different orientations of the magnetic field with respect to the plane of the films were achieved by using a sample probe with a rotator. Resistance was measured in Delta mode using a synchronized set of Keithley 6221 current source and Keithley 2812A nanovoltmeter (*64*). The magnetic field was swept from zero to maximum fields at the rate of 3 T per minute while voltage and current data were collected. Here, we define the upper critical field $H_{c2}$ as the magnetic field at which the resistance increases to 50% of the normal state resistance (*Rn*) just above the critical temperature ($T_c$). Temperature was controlled using a PID controller with input from a capacitive thermometer mounted near the sample stage. This type of thermometer is known for having almost no field dependency, which is ideal for high magnetic field measurements.

The measurements in pulsed magnetic fields were conducted at NHMFL Pulsed Field Facility in Los Alamos National Laboratory. Four-wire electrical resistance measurements were performed in a 65 T short pulse magnet with He-4 cryostat on a probe with *in situ* rotating capabilities(*65*). The sample resistance was measured using pulsed direct current (DC) method based upon a modified form of the approach described in Ref. (*66*). Typical current durations were on the order of ~10 $\mu$s.

**Data presentation**: For each Eu concentration and field orientation, magnetoresistance *R(H)* data were collected at 11-16 temperatures per data set (See Supplemental Information for *R(H)* data). For pulsed field measurement, each *R(H)* data curve was assigned the corresponding thermometer reading at the start of the pulse. For DC field measurement, we used temperature reading from a capacitive thermometer mounted near the sample stage. To facilitate direct comparison across samples and different measurement setup, we represented the data as the colormaps in Fig. 2. The *R(H)* curves were resampled onto a common field grid with 0.25 T steps by selecting the nearest measured point to each grid value. The resulting (*T, B, R*) triplets from all temperatures were combined and interpolated using a natural neighbor interpolation function, producing a smooth resistance surface *R(T,B)*, which is then represented as colormaps as in Figure 2. The upper critical field $H_{c2}(T)$ was extracted for resistance thresholds corresponding to fractions of the normal-state resistance *Rn* by identifying all grid points where *R(T,B)* matched the target within ±0.001 of the target resistance.

The uncertainty from this process is checked by removing individual data points from the input set, re-interpolating *R(T,B)* and comparing the resulting $H_{c2}(T)$ values with the originals This measures the sensitivity of the interpolation to data sparsity and local irregularities in the sampling grid. This type of uncertainty becomes significant in regions where *R(T)* relationship becomes non-linear, i.e. in vicinity of the superconducting transition and near zero resistance and less significant at the middle of the transition where *R(T)* is almost linear and where we sample $H_{c2}(T)$ data.

## X-Ray Diffraction

X-ray diffraction (XRD) of the samples is taken using a rotating anode high-resolution X-ray diffractometer (Rigaku SmartLab). The X-ray energy is fixed at the Cu-Kα energy of 8.04 keV.

## Fourier-Transform Infrared Spectroscopy

We measured reflectivity spectra with a Bruker Vertex 80v Fourier transform spectrometer in quasi-normal incidence geometry. As a source, we used an in-built Hg arc lamp. The total measurement range of ~10 − 800 $cm^{-1}$ was covered by detecting the far-infrared radiation with two different bolometers with working temperatures of 4.2 K and 1.6 K for the ranges above and below ~40 $cm^{-1}$, respectively. The absolute reflectivities (see **Supplemental Information**) were obtained by taking the ratio at a given temperature between the reflected spectrum from the sample (either bare LSAT or NENO/LSAT) and that from a gold reference placed next to it. To this end, the sample holder was mounted on the cold finger of a He flow cryostat, equipped with a motorized stage.

In the case of temperature-dependent normalized reflectivity, to reduce the experimental error arising from changing the sample position, the spectra were measured on a fixed spot on the sample while the temperature was varied (*54*). Then, these were normalized by a reference spectrum taken at 22.5 K. This temperature was chosen to be slightly above 21 K, the onset temperature of superconductivity in NENO, identified by DC transport measurements.

## DFT calculations

All DFT calculations were performed using the Vienna ab initio simulation package (VASP) software (*67*) with ECUT=520eV, and 8×8×6 k-grid sampling for $\sqrt{2}\times\sqrt{2}\times 2$ supercell, and the exchange-correlation effects were treated using the strongly-constrained-and-appropriately-normed (SCAN) meta-GGA functional (*68*). Spin-orbit coupling effects were included self-consistently. We use a $\sqrt{2}\times\sqrt{2}\times 2$ supercell to allow spontaneous spin ordering along in-plane or out-of-plane directions. The ground state obtained in the DFT calculation is a C-AFM magnetic configuration (*18*). We then computed the tight-binding Kohn-Sham Hamiltonian on the maximally localized Wannier basis as extracted from our DFT calculations using the Wannier90 software (*69*). The Wannier basis includes all 3d- and the $4p_z$-orbitals of Ni, all p-orbitals of O, and all f-orbitals of Eu and Nd, which are sufficient to describe the bands near the Fermi level from -4eV to 2eV as well as the dominant magnetic moments (see **Supplemental Information** for details). Next, the Wannier tight-binding model was mapped to a Heisenberg model using the TB2J software (*70*), which directly provides the exchange interactions and allows us to compute the corresponding effective exchange field on the Ni band conduction electrons.

# References

1. D. Li *et al.*, Superconductivity in an infinite-layer nickelate. *Nature* **572**, 624-627 (2019).
2. M. Osada *et al.*, Nickelate superconductivity without rare-earth magnetism: (La, Sr)$NiO_2$. *Advanced Materials* **33**, 2104083 (2021).
3. S. Zeng *et al.*, Superconductivity in infinite-layer nickelate $La_{1-x}Ca_xNiO_2$ thin films. *Science advances* **8**, eabl9927 (2022).
4. M. Osada *et al.*, A superconducting praseodymium nickelate with infinite layer structure. *Nano letters* **20**, 5735-5740 (2020).
5. W. Wei, D. Vu, Z. Zhang, F. J. Walker, C. H. Ahn, Superconducting $Nd_{1-x}Eu_xNiO_2$ thin films using in situ synthesis. *Science Advances* **9**, eadh3327 (2023).
6. S. L. E. Chow, Z. Luo, A. Ariando, Bulk superconductivity near 40 K in hole-doped $SmNiO_2$ at ambient pressure. *Nature* **642**, 58-63 (2025).
7. A. S. Botana, M. R. Norman, Similarities and differences between $LaNiO_2$ and $CaCuO_2$ and implications for superconductivity. *Physical Review X* **10**, 011024 (2020).
8. A. S. Botana, F. Bernardini, A. Cano, Nickelate Superconductors: An Ongoing Dialog between Theory and Experiments. *Journal of Experimental and Theoretical Physics* **132**, 618-627 (2021).
9. J. Karp *et al.*, Many-Body Electronic Structure of $NdNiO_2$ and $CaCuO_2$. *Physical Review X* **10**, 021061 (2020).
10. B. H. Goodge *et al.*, Doping evolution of the Mott–Hubbard landscape in infinite-layer nickelates. *Proceedings of the National Academy of Sciences* **118**, e2007683118 (2021).
11. J. F. Mitchell, A Nickelate Renaissance. *Frontiers in Physics* **9**, 813483 (2021).
12. J. P. Carbotte, E. Schachinger, D. N. Basov, Coupling strength of charge carriers to spin fluctuations in high-temperature superconductors. *Nature* **401**, 354-356 (1999).
13. R. Cervasio *et al.*, Optical Properties of Superconducting $Nd_{0.8}Sr_{0.2}NiO_2$ Nickelate. *ACS Applied Electronic Materials* **5**, 4770-4777 (2023).
14. B. Cheng *et al.*, Evidence for d-wave superconductivity of infinite-layer nickelates from low-energy electrodynamics. *Nature Materials* **23**, 775-781 (2024).
15. K. W. Lee, W. E. Pickett, Infinite-layer $LaNiO_2$: $Ni^{1+}$ is not $Cu^{2+}$. *Physical Review B* **70**, 165109 (2004).
16. Y. Zhang, J. Zhang, X. He, J. Wang, P. Ghosez, Rare-earth control of phase transitions in infinite-layer nickelates. *PNAS Nexus* **2**, pgad108 (2023).
17. P. Adhikary, S. Bandyopadhyay, T. Das, I. Dasgupta, T. Saha-Dasgupta, Orbital-selective superconductivity in a two-band model of infinite-layer nickelates. *Physical Review B* **102**, 100501 (2020).
18. R. Zhang *et al.*, Magnetic and f-electron effects in $LaNiO_2$ and $NdNiO_2$ nickelates with cuprate-like $3d_{x2-y2}$ band. *Communications Physics* **4**, 118 (2021).
19. K. Foyevtsova, I. Elfimov, G. A. Sawatzky, Distinct electridelike nature of infinite-layer nickelates and the resulting theoretical challenges to calculate their electronic structure. *Physical Review B* **108**, 205124 (2023).
20. W. Sun *et al.*, Evidence for Anisotropic Superconductivity Beyond Pauli Limit in Infinite-Layer Lanthanum Nickelates. *Advanced Materials* **35**, 2303400 (2023).
21. B. Y. Wang *et al.*, Effects of rare-earth magnetism on the superconducting upper critical field in infinite-layer nickelates. *Science Advances* **9**, eadf6655 (2023).
22. L. Chow *et al.*, Pauli-limit violation in lanthanide infinite-layer nickelate superconductors. *arXiv preprint arXiv:2204.12606*, (2022).
23. B. Y. Wang *et al.*, Isotropic Pauli-limited superconductivity in the infinite-layer nickelate $Nd_{0.775}Sr_{0.225}NiO_2$. *Nature Physics* **17**, 473-477 (2021).
24. L. E. Chow, A. Ariando, Infinite-layer nickelate superconductors: A current experimental perspective of the crystal and electronic structures. *Frontiers in Physics* **10**, 834658 (2022).

25. L. E. Chow *et al.*, Pairing symmetry in infinite-layer nickelate superconductor. *arXiv preprint arXiv:2201.10038*, (2022).
26. S. P. Harvey *et al.*, Evidence for nodal superconductivity in infinite-layer nickelates. *arXiv preprint arXiv:2201.12971*, (2022).
27. Q. Gu *et al.*, Single particle tunneling spectrum of superconducting $Nd_{1-x}Sr_xNiO_2$ thin films. *Nature Communications* **11**, 6027 (2020).
28. M. Tinkham, *Introduction to superconductivity*. (Courier Corporation, 2004), vol. 1.
29. G. Blatter, M. V. Feigel'man, V. B. Geshkenbein, A. I. Larkin, V. M. Vinokur, Vortices in high-temperature superconductors. *Reviews of Modern Physics* **66**, 1125-1388 (1994).
30. L. I. Glazman, A. E. Koshelev, Thermal fluctuations and phase transitions in the vortex state of a layered superconductor. *Physical Review B* **43**, 2835-2843 (1991).
31. Y. Cao, J. M. Park, K. Watanabe, T. Taniguchi, P. Jarillo-Herrero, Pauli-limit violation and re-entrant superconductivity in moiré graphene. *Nature* **595**, 526-531 (2021).
32. Z. Wu *et al.*, Enhanced triplet superconductivity in next generation ultraclean UTe2. *Proceedings of the National Academy of Sciences* **121**, e2403067121 (2024).
33. T. Shishidou, H. G. Suh, P. M. R. Brydon, M. Weinert, D. F. Agterberg, Topological band and superconductivity in $UTe_2$. *Physical Review B* **103**, 104504 (2021).
34. J. Ishizuka, S. Sumita, A. Daido, Y. Yanase, Insulator-Metal Transition and Topological Superconductivity in $UTe_2$ from a First-Principles Calculation. *Physical Review Letters* **123**, 217001 (2019).
35. Y. Xu, Y. Sheng, Y.-f. Yang, Quasi-Two-Dimensional Fermi Surfaces and Unitary Spin-Triplet Pairing in the Heavy Fermion Superconductor $UTe_2$. *Physical Review Letters* **123**, 217002 (2019).
36. S. Ran *et al.*, Extreme magnetic field-boosted superconductivity. *Nature Physics* **15**, 1250-1254 (2019).
37. S. Uji *et al.*, Magnetic-field-induced superconductivity in a two-dimensional organic conductor. *Nature* **410**, 908-910 (2001).
38. H. W. Meul *et al.*, Observation of Magnetic-Field-Induced Superconductivity. *Physical Review Letters* **53**, 497-500 (1984).
39. M. Giroud *et al.*, Magnetic field-induced superconductivity in the ferromagnetic state of $HoMo_6S_8$. *Journal of low temperature physics* **69**, 419-450 (1987).
40. V. Jaccarino, M. Peter, Ultra-High-Field Superconductivity. *Physical Review Letters* **9**, 290-292 (1962).
41. O. H. Fischer, Properties of high-field superconductors containing localized magnetic moments. *Helv. Phys. Acta* **45**, 331-397 (1972).
42. K.-i. Hiraki *et al.*, $^{77}$Se NMR Evidence for the Jaccarino–Peter Mechanism in the Field Induced Superconductor, $\lambda$-$(BETS)_2FeCl_4$. *Journal of the Physical Society of Japan* **76**, 124708 (2007).
43. H. W. Meul *et al.*, Superconductivity induced by a magnetic field. *Physica B+C* **126**, 44-50 (1984).
44. E. Helfand, N. Werthamer, Temperature and purity dependence of the superconducting critical field, H c 2. *Physical Review Letters* **13**, 686 (1964).
45. E. Helfand, N. Werthamer, Temperature and purity dependence of the superconducting critical field, H c 2. II. *Physical Review* **147**, 288 (1966).
46. N. Werthamer, E. Helfand, P. Hohenberg, Temperature and purity dependence of the superconducting critical field, H c 2. III. Electron spin and spin-orbit effects. *Physical Review* **147**, 295 (1966).
47. N. Werthamer, W. McMillan, Temperature and Purity Dependence of the Superconducting Critical Field H c 2. IV. Strong-Coupling Effects. *Physical Review* **158**, 415 (1967).
48. W. Wei *et al.*, Large upper critical fields and dimensionality crossover of superconductivity in the infinite-layer nickelate $La_{0.8}Sr_{0.2}NiO_2$. *Physical Review B* **107**, L220503 (2023).
49. A. Gurevich *et al.*, Very high upper critical fields in$MgB_2$ produced by selective tuning of impurity scattering. *Superconductor Science and Technology* **17**, 278 (2003).

50. H. Ji *et al.*, Rotational symmetry breaking in superconducting nickelate $Nd_{0.8}Sr_{0.2}NiO_2$ films. *Nature Communications* **14**, 7155 (2023).
51. I. M. Vishik *et al.*, Phase competition in trisected superconducting dome. *Proceedings of the National Academy of Sciences* **109**, 18332-18337 (2012).
52. Y. He *et al.*, Rapid change of superconductivity and electron-phonon coupling through critical doping in Bi-2212. *Science* **362**, 62-65 (2018).
53. J. A. Sobota, Y. He, Z.-X. Shen, Angle-resolved photoemission studies of quantum materials. *Reviews of Modern Physics* **93**, 025006 (2021).
54. N. M. Rugheimer, A. Lehoczky, C. V. Briscoe, Microwave Transmission- and Reflection-Coefficient Ratios of Thin Superconducting Films. *Physical Review* **154**, 414-421 (1967).
55. D. C. Mattis, J. Bardeen, Theory of the Anomalous Skin Effect in Normal and Superconducting Metals. *Physical Review* **111**, 412-417 (1958).
56. Z. Tagay *et al.*, BCS d-wave behavior in the terahertz electrodynamic response of electron-doped cuprate superconductors. *Physical Review B* **104**, 064501 (2021).
57. D. A. G. Bruggeman, Berechnung verschiedener physikalischer Konstanten von heterogenen Substanzen. I. Dielektrizitätskonstanten und Leitfähigkeiten der Mischkörper aus isotropen Substanzen. *Annalen der Physik* **416**, 636-664 (1935).
58. Y. Yu, S. Iskakov, E. Gull, K. Held, F. Krien, Pairing boost from enhanced spin-fermion coupling in the pseudogap regime. *Physical Review B* **112**, L041105 (2025).
59. S. L. E. Chow, A. Ariando, Nickel Age of High-Temperature Superconductivity. *Advanced Materials Interfaces* **12**, 2400717 (2025).
60. X. Yan *et al.*, Superconductivity in an ultrathin multilayer nickelate. *Science Advances* **11**, eado4572 (2025).
61. G. A. Pan *et al.*, Superconductivity in a quintuple-layer square-planar nickelate. *Nature Materials* **21**, 160-164 (2022).
62. S. Zeng *et al.*, Phase Diagram and Superconducting Dome of Infinite-Layer $Nd_{1-x}Sr_xNiO_2$ Thin Films. *Physical Review Letters* **125**, 147003 (2020).
63. W. Wei *et al.*, Solid state reduction of nickelate thin films. *Physical Review Materials* **7**, 013802 (2023).
64. A. V. Suslov, Stand alone experimental setup for DC transport measurements. *Review of Scientific Instruments* **81**, 075111 (2010).
65. X. Willis, X. Ding, J. Singleton, F. F. Balakirev, Cryogenic goniometer for measurements in pulsed magnetic fields fabricated via additive manufacturing technique. *Review of Scientific Instruments* **91**, 036102 (2020).
66. M. Leroux *et al.*, Dynamics and Critical Currents in Fast Superconducting Vortices at High pulsed Magnetic Fields. *Physical Review Applied* **11**, 054005 (2019).
67. G. Kresse, J. Furthmüller, Efficiency of ab-initio total energy calculations for metals and semiconductors using a plane-wave basis set. *Computational Materials Science* **6**, 15-50 (1996).
68. J. Sun, A. Ruzsinszky, J. P. Perdew, Strongly Constrained and Appropriately Normed Semilocal Density Functional. *Physical Review Letters* **115**, 036402 (2015).
69. N. Marzari, D. Vanderbilt, Maximally localized generalized Wannier functions for composite energy bands. *Physical Review B* **56**, 12847-12865 (1997).
70. X. He, N. Helbig, M. J. Verstraete, E. Bousquet, TB2J: A python package for computing magnetic interaction parameters. *Computer Physics Communications* **264**, 107938 (2021).

# Figures

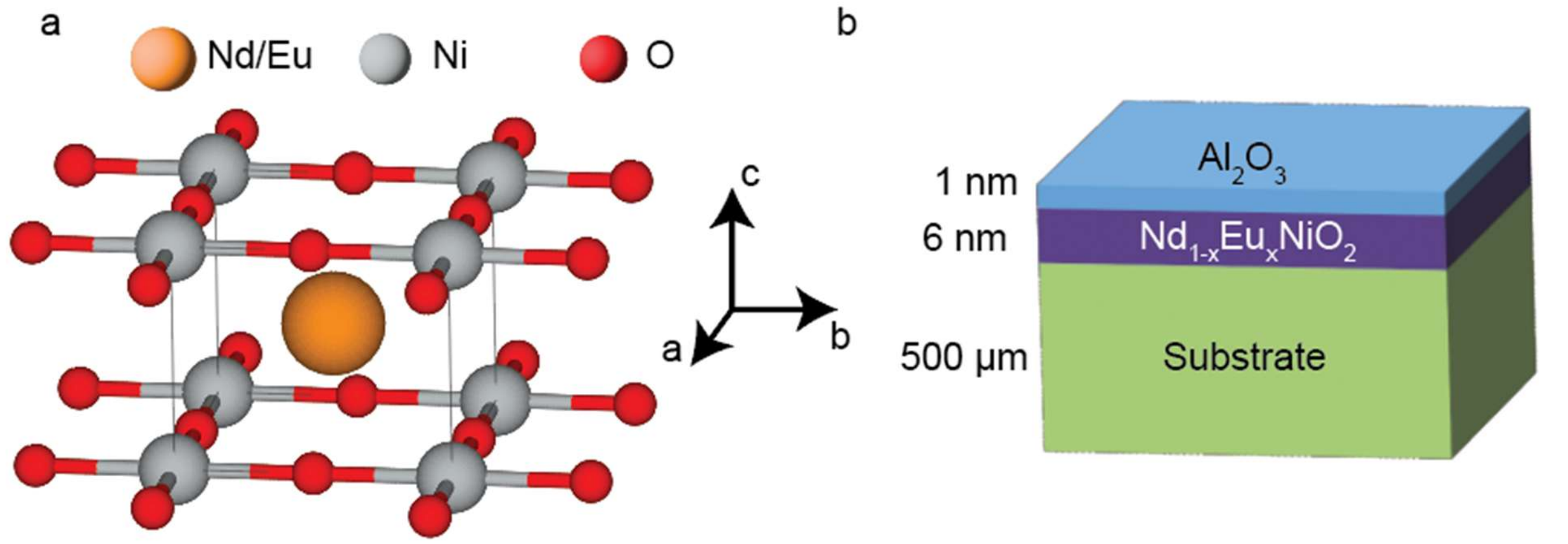

**Figure 1. Thin-film infinite-layer nickelates. (a)** Schematic showing the atomic arrangement in Eu doped $NdNiO_2$. The rare-earth layer (Nd/Eu) is intercalated between the $NiO_2$ planes in an infinite layer structure **(b)** Schematic of the NENO thin-film sample. NENO thin film of 6-nm-thickness was grown on a 500-µm-thick LSAT substrate. The film is capped with a 1nm thick $Al_2O_3$ layer.

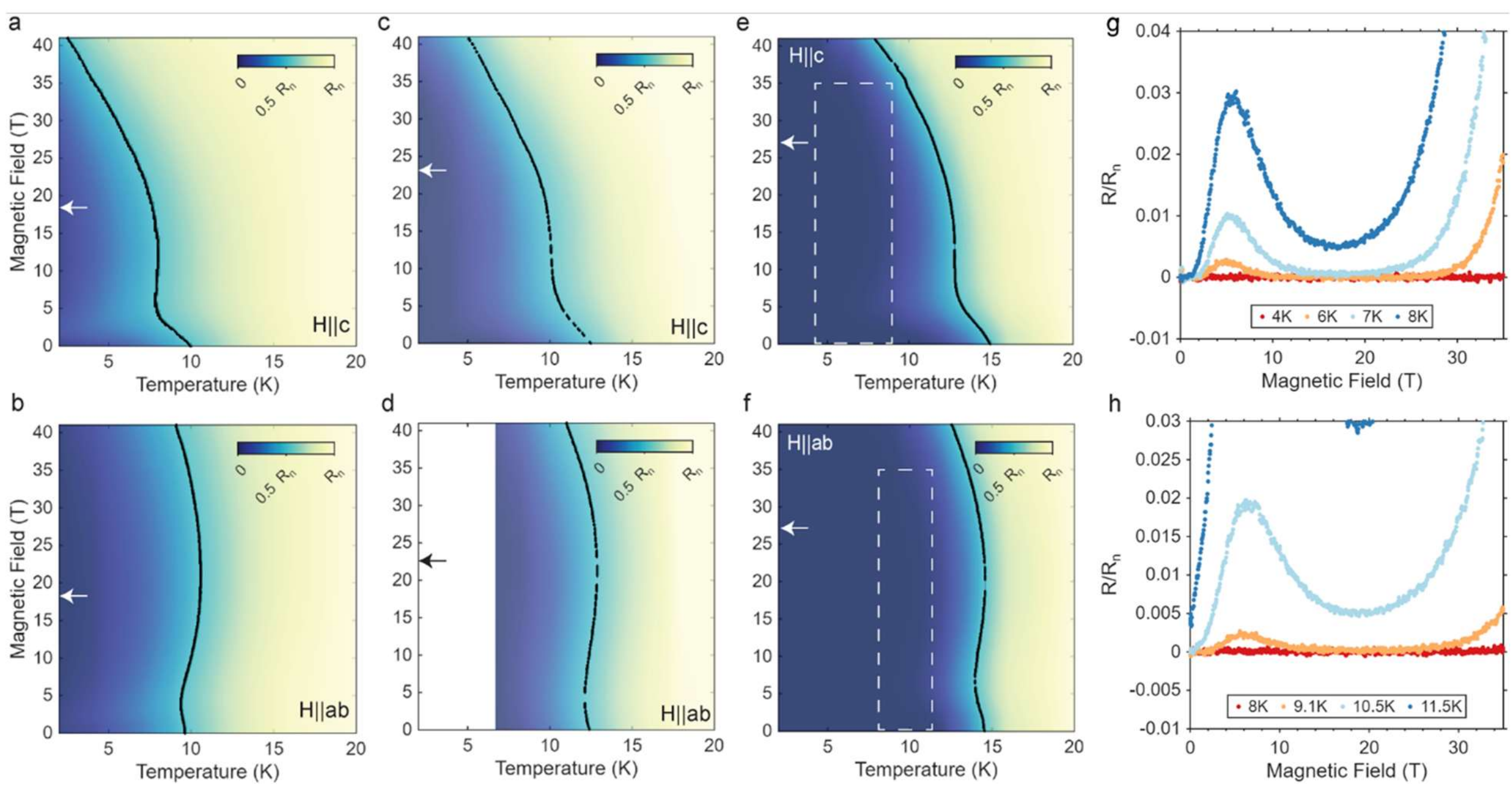

**Figure 2. Magneto transport measurements of thin films. (a-f)** Color plots represent normalized resistance *R/Rn(T, H)* and $H_{c2}(T)$ data of x =0.20 (panel **a, b**), 0.22 (panel **c, d**), and 0.35 (panel **e, f**) $Nd_{1-x}Eu_xNiO_2$ samples from left to right in out-of-plane magnetic field **(a, c, e)** and in-plane magnetic field **(b, d, f)**. Colors represent normalized resistance *R/Rn* with scale indicated by color bar. Contour lines indicate 50% *Rn* resistances, corresponding to the temperature dependence of $H_{c2}$ with respect to $T_{c,50\%Rn}$ criterion. Arrows on vertical axes indicate the nominal Pauli limit of upper critical field, calculated by $H_{c2}^{P}$=1.86×$T_{c,50\%Rn}$. Boxes in dashed lines in panels **(e)** and **(f)** represent regions where MR data in **(g)** and **(h)** are depicted. **(g, h)** Zero resistances is induced at moderate temperatures by a large magnetic field in both **H**||c and **H**||ab, respectively, of the x=0.35 sample.

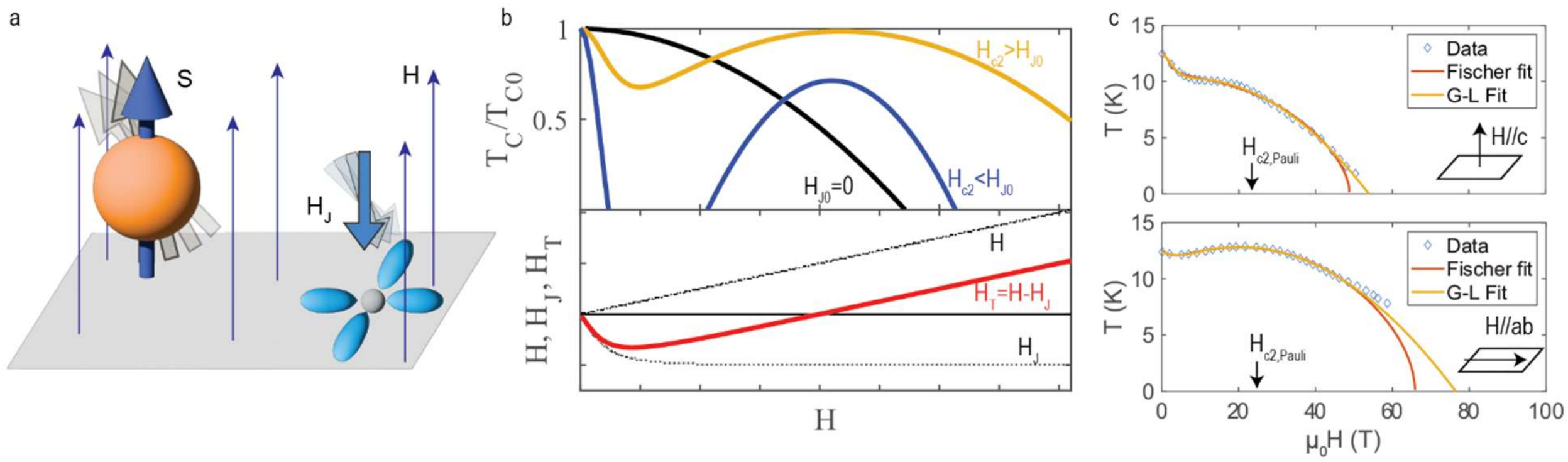

**Figure 3**. **Magnetism and superconductivity in the nickelates. (a)** Schematic of the Jaccarino-Peter effect showing local magnetic moment **S** with an exchange field $\mathbf{H_J}$ acting on electron orbitals and being anti-parallel to **S**. In a polarizing magnetic field **H**, **S** aligns to **H** and causes $\mathbf{H_J}$ to oppose **H**, reducing the total magnetic field on electron orbitals **(b)** Bottom panel: in compensation effect, the exchange field $\mathbf{H_J}$, created by AFM coupling, opposes the external field **H**, leading to a total magnetic field $\mathbf{H_T}$=**H**+$\mathbf{H_J}$. In paramagnetic case, |$\mathbf{H_T}$|=$H$-$H_{J0}B_J(T, H)$, where $H_{J0}$ is the saturated exchange field when the local moments are polarized and $B_J(T, H)$ is defined by the Brillouin function. Top panel: different $H_{c2}(T_c/T_{c0})$ of superconductors with $H_{J0}$=0 (black), $H_{J0}$>$H_{c2}$ (blue) and $H_{J0}$<$H_{c2}$ (yellow). **(c)** $H_{c2}(T)$ data of the sample with x=0.22 (diamonds) fitted to the Fischer formula (red solid curve) and the Ginzburg-Landau (G-L) approach (yellow solid curve), indicating compatibility with the Jaccarino-Peter effect. At low temperatures, both the experimental and the G-L fitting curve lay above the Fischer fitting curve. The measured $H_{c2}$ significantly exceeds the Pauli limit even at temperatures far above 0K.

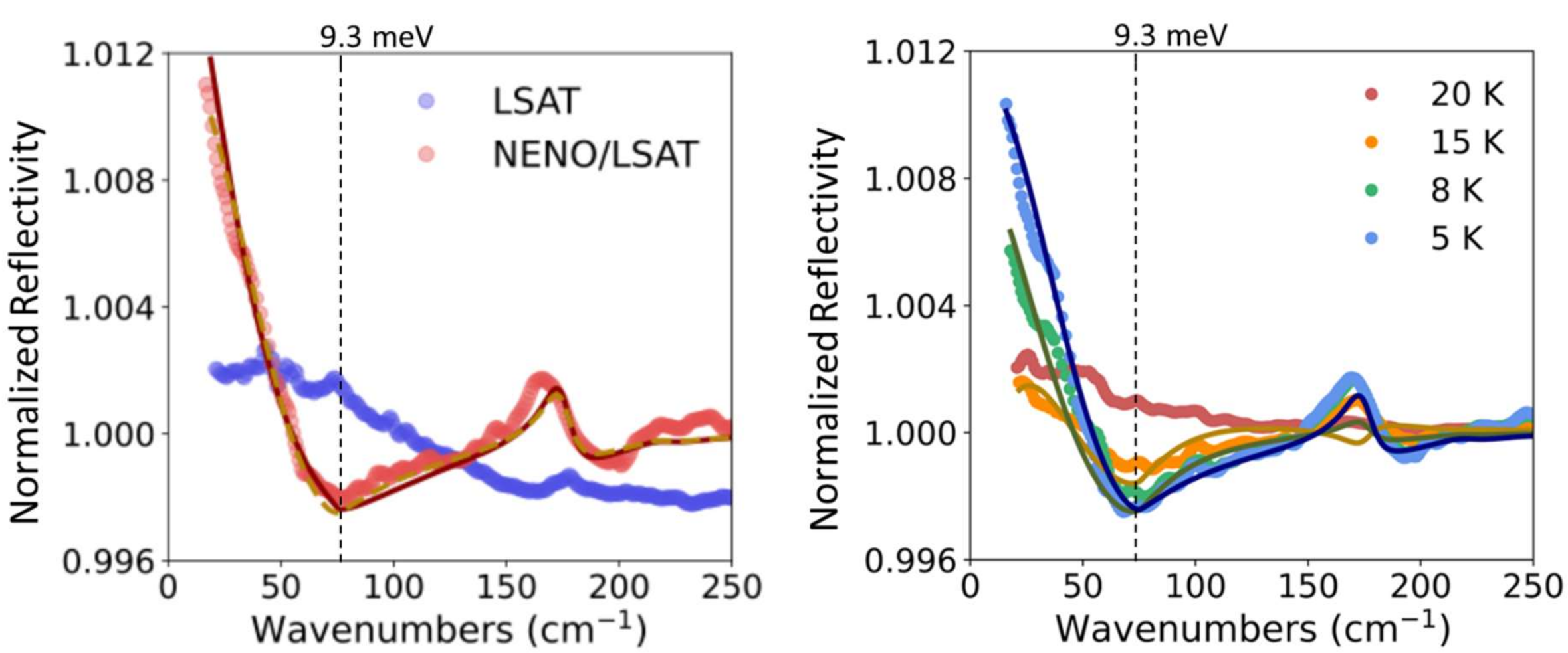

**Figure 4. Evidence of unconventional pairing from optical measurements of the superconducting gap. (a)** Reflectivity of the LSAT substrate (blue circles) and the NENO/LSAT film (red circles) measured at $T$ = 3.5 K, normalized by the same quantity measured at 22.5 K (*i.e.*, above the NENO superconducting $T_C$). The solid line is a multilayer fit (see main text) in which the superconducting response of NENO was modeled by an $s$-wave Mattis-Bardeen model. The dashed line corresponds to a $d$-wave Mattis-Bardeen fit in the presence of residual uncondensed quasiparticles. The gap size was set to 75 cm$^{-1}$ for both fits. The extracted normal state conductivity was 100 $\Omega^{-1}\text{cm}^{-1}$ and 500 $\Omega^{-1}\text{cm}^{-1}$ for *s*- and *d*-wave fits, respectively. Note that in both fits the substrate phonon parameters were kept fixed, so the contribution at ~170 $\text{cm}^{-1}$ is mainly attributable to a change in multilayer reflectivity at the LSAT phonon frequency due to the opening of the superconducting gap in NENO. **(b)** Temperature dependence of the normalized reflectivity of NENO/LSAT (colored circles). The fits were calculated with the same *d*-wave Mattis-Bardeen model as the dashed line in (a). The gap was kept fixed, while the contribution of uncondensed quasiparticles was allowed to increase with temperature (see also **Supplemental Information**).